\documentclass[prb,twocolumn,showpacs,preprintnumbers,amsmath,amssymb,superscriptaddress]{revtex4-1}
\pdfoutput=1 
\usepackage{gensymb}
\usepackage{graphicx} 
\usepackage{amsmath} 
\usepackage{placeins}
\usepackage{booktabs}
\usepackage{multirow}
\usepackage{caption}
\usepackage{subcaption}
\usepackage{textcomp}
\usepackage{hyperref}
\usepackage{xcolor}

\begin{document}

\title{The magnetic structure factor of correlated moments in small-angle neutron scattering}

\author{D. Honecker}
\email{dirk.honecker@uni.lu}
\affiliation{Institut Laue-Langevin, 38042 Grenoble, France}
\altaffiliation{now at: University of Luxembourg, L-1511 Luxembourg, Grand Duchy of Luxembourg, Luxembourg}

\author{L. \surname{Fern\'{a}ndez Barqu\'{i}n}}
\affiliation{Departamento CITIMAC, Faculty of Science, University of Cantabria, 39005 Santander, Spain}

\author{P. Bender}
\affiliation{University of Luxembourg, 1511 Luxembourg, Grand Duchy of Luxembourg, Luxembourg}

\date{\today}

\begin{abstract}
The interplay between structural and magnetic properties of nanostructured magnetic materials allows to realize unconventional magnetic effects, which results in a demand for experimental techniques to determine the magnetization profile with nanoscale resolution.
Magnetic small-angle neutron scattering (SANS) probes both the chemical and magnetic nanostructure and is thus a powerful technique e.g. for the characterization of magnetic nanoparticles.
Here, we show that the conventionally used particle-matrix approach to describe SANS of magnetic particle assemblies, however, leads to a flawed interpretation. 
As remedy, we provide general expressions for the field-dependent 2D magnetic SANS cross-section of correlated moments.
It is shown that for structurally disordered ensembles the magnetic structure factor is in general, and contrary to common assumptions, (i)  anisotropic also in zero field, and (ii) that even in saturation the magnetic structure factor deviates from the nuclear one. 
These theoretical predictions explain qualitatively the intriguing experimental, polarized SANS data of an ensemble of dipolar-coupled iron oxide nanoparticles.
\end{abstract}


\maketitle

\section{Introduction}
A key challenge in magnetism remains the visualization of complex three dimensional (3D) magnetization vector fields\cite{Pacheco2017}. 
Individual magnetic structures can be characterized with advanced imaging techniques like X-ray nanotomography, which allows to reveal the internal 3D magnetization profile\cite{Donnelly2017}, however, with a resolution limit of around 50 nm and a penetration depth in the micrometer range.
In contrast, neutrons allow to study the magnetism in bulk samples, with magnetic small-angle neutron scattering (SANS) being a powerful tool to probe the spatial magnetization distribution of nanostructured systems. 
Magnetic SANS measures the diffuse, forward scattering arising from nanoscale variations of the magnitude and orientation of the local magnetization vector $\mathbf{M}(\mathbf{r})$  from its mean value, reflecting the underlying magnetic microstructure, which depends on magnetic interactions, structural features (e.g., particle-size distribution and crystallographic orientations), and on the applied magnetic field. Examples cover nanocrystalline ferromagnets and steels (see e.g., Ref.~\onlinecite{RevModPhys.91.015004} and references therein), nanowire arrays \cite{Grigoryev2007,maurer2014ordered,grutter2017complex}, spin glasses \cite{Calderon2005, Giot2008}, ferromagnetic clusters in  alloys \cite{Bhatti2012, Laver2010}, magnetic recording media \cite{Wismayer2006}, nanogranular films \cite{bellouard1996magnetic,farrell2006small,majetich2006magnetostatic,alba2016magnetic}, as well as nanoparticles \cite{yusuf2006experimental,krycka2010core,Disch2012,Bender2019} and their assemblies in the 3D bulk \cite{Sachan2008,Ridier2017, Bender2019,Bersweiler2019} or in liquids \cite{Meriquet2006,klokkenburg2007dipolar,heinemann2007reordering,avdeev2010small,Barrett2011,Rajnak2015,Fu2016,Mertelji2019}. 

For nanostructured bulk ferromagnets subjected to non-saturating magnetic fields, it has been shown that the appearance of long-wavelength magnetization fluctuations lead to a strong variation of the scattering cross-section, not only in scattering intensity, but also in the observed scattering anisotropy and momentum transfer behavior\cite{michels2014Review}. The field-dependent magnetic scattering is associated to the so-called spin-misalignment scattering, which can exceed the residual magnetic scattering measured at absolute saturation by several orders of magnitude. 

Micromagnetism is a destined theory to describe the mesoscale magnetization vector field, which is probed by magnetic SANS. Using a micromagnetic approach allows to analyze the field-dependent SANS cross-section of nanostructured ferromagnets, i.e., bulk material which consists of ferromagnetic nanoparticles that are embedded in and exchange coupled to a ferromagnetic matrix, and to extract relevant magnetic material parameters, like exchange coupling as well as magnetic anisotropy and magnetostatic fields \cite{michels2014Review}. Analytical solutions of the micromagnetic equations are strictly speaking only valid in the approach to saturation, i.e., in the single domain state of a bulk ferromagnet, where the longitudinal magnetization is assumed to be hardly affected by the magnetic field. Noteworthy, the micromagnetic description reveals that two transversal magnetization components along and perpendicular to the beam exhibit  different functional forms and hence distinct scattering contributions due to the effect of dipolar interactions. This is illustratively shown by micromagnetic simulations of magnetic SANS, which allows to decrypt the individual magnetic Fourier components \cite{michels2014MagSim}.

Most often however, magnetic SANS is interpreted in terms of a particle-matrix approach for spatially localised 3D objects, which adopts the formalism used in nuclear scattering. Regarding magnetic nanoparticles, it is usually assumed that they are homogeneously magnetized (i.e., each single-domain particle acts as a dipole with a net moment $\boldsymbol{\mu}$) and thus that the (magnetic) cross-sections can be described by a particle form factor amplitude $F(q)$ weighted with a sum of Langevin functions to consider the field variation of the particle moment orientation \cite{wiedenmann2005}. This description is a good approximation for dilute systems and if magnetocrystalline anisotropy is negligible, e.g., for a Brownian ensemble of mobile particles. Effects of the magnetocrystalline anisotropy on the magnetic SANS cross-section are expected for superparamagnetic particles embedded in a solid nonmagnetic matrix \cite{Michels2002}. For non-dilute, mobile magnetic particle systems, the structural arrangement of particles and hence the (nuclear) interparticle correlations are determined by the magnetic dipolar, van-der-Waals and repulsive (electrostatic or steric) interactions as well as particle shape, which eventually leads to the occurrence of self-assembled aggregates \cite{klokkenburg2007dipolar,Disch2013,Fu2016,Mertelji2019}. 

In particular, for measurements at zero and high, saturating field strength, which are most commonly performed,  it is for simplicity assumed that the magnetic moments of the particles are either totally uncorrelated, such that a  structure factor is negligible in the magnetic scattering, \cite{Gazeau2003} or fully aligned and thus the nuclear and magnetic structure factor are taken as equal,  given by the spatial particle arrangement and the concomitant pair distance distribution function
\cite{Hayter1991,wiedenmann2006dynamics,yusuf2006experimental,klokkenburg2007dipolar,heinemann2007reordering,avdeev2010small}.
The effect of magnetic correlations between particle moments and the resulting magnetic scattering interference is scarcely addressed in magnetic field dependent SANS, which is however of relevance for systems like dense powders or other (more ordered) bulk structures, e.g.,  mesocrystals and nanowire arrays.

In magnetic neutron diffraction, so-called spin-pair correlation functions have been introduced, which allow to describe the magnetic diffuse scattering of  (short-range) coupled atomic magnetic moments\cite{blech1964spin,paddison2013spinvert,frandsen2015magnetic}. Here, we extend the approach to describe magnetic SANS, e.g., of cooperative superparamagnets exhibiting a local order of correlated magnetic particle moments. The approach is inspired by a decomposition of scattering components performed in X-ray cross-correlation analysis to provide information on the local structural arrangement in a disordered system \cite{Altarelli2010}. The theoretical framework which we derive here has important consequences on the interpretation of diffuse magnetic neutron scattering from dense, locally ordered magnetic entities (atoms or particles) observed on a 2D detector array. The findings of the directional dependence of the magnetic structure factor should be also taken into account for  x-ray resonant scattering of magnetic nanostructures\cite{Kortright2005,Bagschik2016,Rackham2019}.

In the following, we will first present the experimental, magnetic SANS results of a powder of interacting single-domain iron oxide nanoparticles.
Afterwards, we derive theoretical expressions for the magnetic structure factor in SANS of correlated moment ensembles, which can qualitatively describe our experimental data.

\begin{figure}[b]
\centering
\includegraphics[width=1\columnwidth]{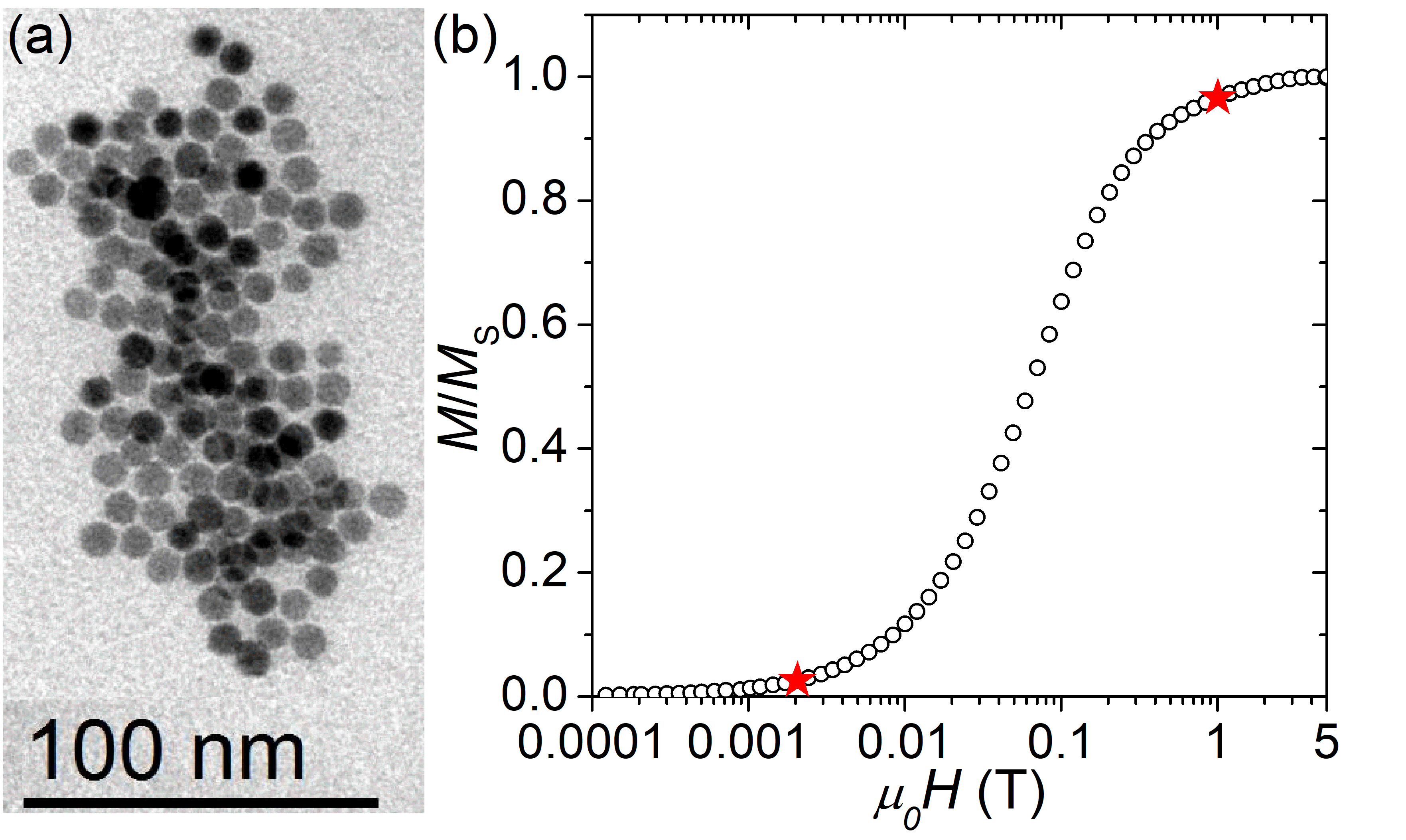}
\caption{\label{Fig1}(a) TEM image of the iron oxide particles. (b) Isothermal magnetization curve $M(H)$ of the sample at 300\,K, normalized to the saturation magnetization $M_{\rm S}$. \textit{Red stars:} Fields at which the POLARIS measurement was performed ($\mu_0H=2\,\mathrm{mT}$, $M/M_{\rm S}=0.026$ and $\mu_0H=1\,\mathrm{T}$, $M/M_{\rm S}=0.966$).}
\end{figure}

\begin{figure*}[ht]
\centering
\includegraphics[width=2\columnwidth]{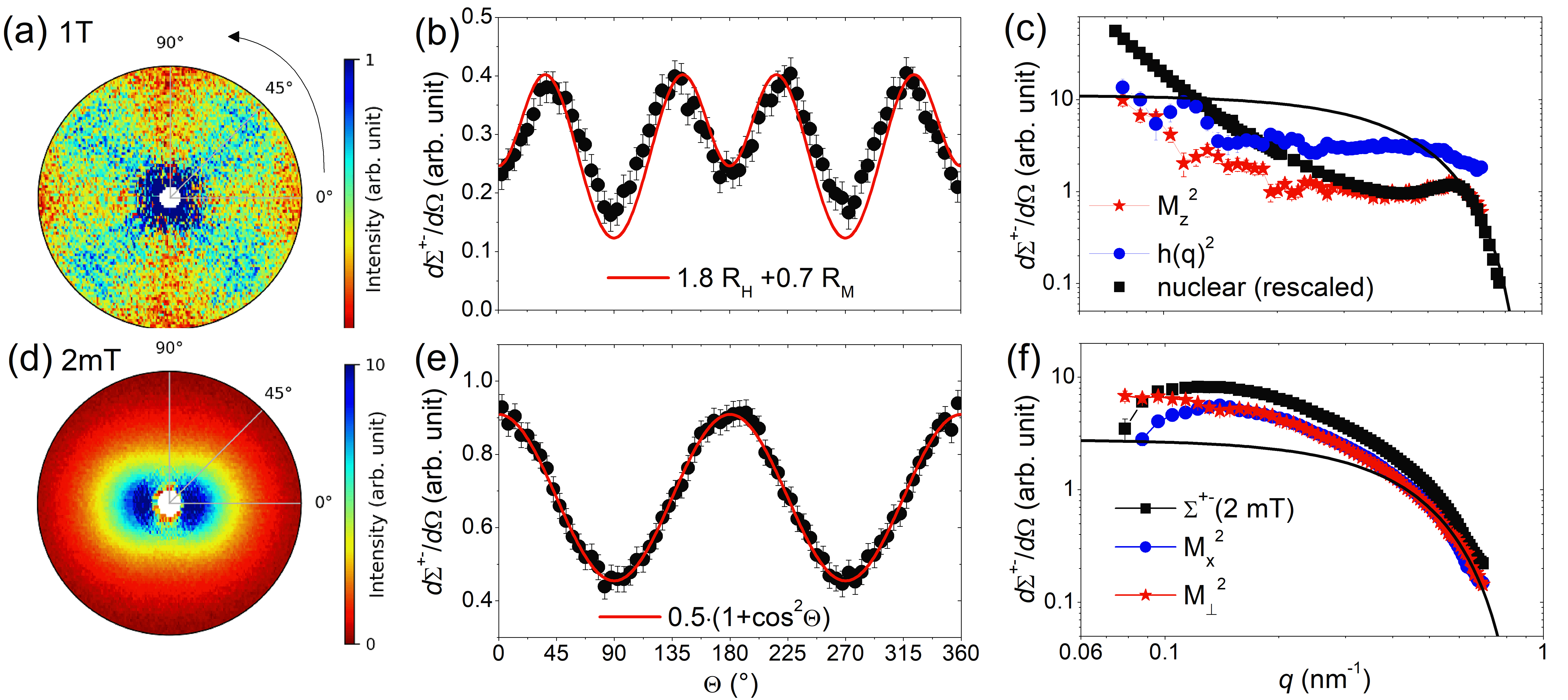}
\caption{\label{Fig2}Experimental results for the spin-flip cross-sections of the nanoparticle powder.
(a) Polar plot of $d \Sigma^{+ -} / d \Omega(q,\Theta)$ at 1\,T ($q=0.07-0.7\,\mathrm{nm^{-1}}$; $\mathbf{H}$ along $\Theta=0^{\circ}$).
(b) Radial average $d \Sigma^{+ -} / d \Omega(\Theta)$ (integrated over $q=0.4-0.7\,\mathrm{nm^{-1}}$), the solid red line corresponds to the fit result using the micromagnetic model (Eq.~\ref{sigmasmperpsf}).
(c) Best fit results using Eq.~\ref{sigmasmperpsf} for the Fourier transform of  the random-perturbation field $h^2(q)$, of the longitudinal magnetization $\widetilde{M}_z(q)^2$ and the purely nuclear scattering intensity $|\widetilde{N}(q)|^2$ (from the non-spin-flip scattering cross-section). For comparison, the solid line displays the form factor of a sphere with radius 10 nm.
 (d) Polar plot of $d \Sigma^{+ -} / d \Omega(q,\Theta)$ in linear scale detected at 2\,mT, 
(e) radial average $d \Sigma^{+ -} / d \Omega(\Theta)$. The solid red line is the theoretical prediction for a random distribution of the particle moments (see text).
(f) Azimuthal average $d \Sigma^{+ -} / d \Omega(q)$, the magnetization component along the beam $\widetilde{M}_x^2$ and the magnetization component in the detector plane $\widetilde{M}_{\perp}^2$ (determined by a fit with $d \Sigma^{+ -} / d \Omega(\mathbf{q})=\widetilde{M}_x^2+\widetilde{M}_{\perp}^2\cos^2\Theta$) .}
\end{figure*}

\section{Experimental Results}
The investigated sample was a powder of maghemite particles coated with dimercatosuccinic acid. Details regarding the synthesis and characterization of the iron oxide particles can be found in Ref.~\onlinecite{bender2018dipolar}. A representative transmission electron microscopy (TEM) image of the particles is shown in Fig.~\ref{Fig1}(a). 
The particles are spherically shaped and nearly monodisperse with a mean core size of $\left\langle D_\mathrm{TEM}\right\rangle=9.7\,\mathrm{nm}$. 
Fig.~\ref{Fig1}(b) displays the isothermal magnetization curve $M(H)$, normalized to the saturation magnetization $M_{\rm S}$ (which is around $330 \,\mathrm{kA/m}$ at 300\,K) of the particle powder. 
According to integral magnetization measurements and M{\"o}ssbauer spectroscopy, the clustered iron-oxide nanoparticles show at room temperature superparamagnetic behavior (vanishing magnetization at zero field) but with signs for cooperative magnetic correlations \cite{bender2018dipolar}. 
This finding is an indication that the superparamagnetic relaxations (i.e., N\'eel relaxation)  of the particle moments is slowed down by dipole-dipole interactions \cite{Dormann1999,Oscar2004,azeggagh2007,dejardin2011}. 
The magnetic characteristics of the ensembles hence significantly depend on local the magnetostatic stray fields felt by the clustered nanoparticles \cite{landi2017,Kuznetsov2018,Trisnanto2019,Fabris2019} and reflects the subtle interplay between dipolar interactions, local magnetic anisotropy and sample homogeneity \cite{Sanchez2020}.

In the last years and with the advent of efficient $^3$He spin filters,  spin-resolved, longitudinal neutron-spin analysis in SANS (POLARIS) \cite{michelsnn2011} is routinely performed to study assemblies of magnetic nanoparticles \cite{wiedenmann2005,Laver2010,krycka2010core,grutter2017complex,Bender2019,bender2018dipolar,orue2018configuration}. The technique allows to separate coherent nuclear and magnetic scattering and provides access to the three-dimensional spatial distribution of magnetic moments. Here, in particular the  two-dimensional (2D) spin-flip (sf) scattering cross-section $d \Sigma^{+ -} / d \Omega$ is of interest as it  exclusively contains magnetic scattering contributions.
The POLARIS experiment was conducted at $300\,\mathrm{K}$ at the instrument D33 at the Institut Laue-Langevin, France \cite{ILLproposal}. Neutron spin-leakage correction was performed with \textsc{GRASP} \cite{GRASP}. We used a mean wavelength of $\lambda=0.6\,\mathrm{nm}$ ($\Delta\lambda/\lambda\approx 10\,\%$) and two different detector distances (13.4\,m and 3\,m), yielding a  $q$-range of $0.07-0.77\,\mathrm{nm^{-1}}$. The horizontal magnetic field $\mathbf{H}\parallel\mathbf{e}_z$ at the sample position was applied  perpendicular to the wave vector $\mathbf{k}_0\parallel\mathbf{e}_x$ of the incident neutron beam. 
The sf scattering cross-section (for $\mathbf{H}\perp \mathbf{k}_0$, with $\mathbf{H}|| \mathbf{e}_z$, $\mathbf{k}_0|| \mathbf{e}_x$) is given as:
\begin{equation}
\label{Eq1}
\begin{aligned}
\frac{d \Sigma^{\pm \mp}}{  d \Omega}(\mathbf{q}) &= \frac{8 \pi^3}{V} \, b_{\rm H}^2 \left( \widetilde{M}_x^2 + \widetilde{M}_y^2 \cos^4\Theta + \widetilde{M}_z^2 \sin^2\Theta \cos^2\Theta \right. \\
  & \left. - (\widetilde{M}_y \widetilde{M}_z^{\ast} + \widetilde{M}_y^{\ast} \widetilde{M}_z) \sin\Theta \cos^3\Theta \right) \,. 
\end{aligned}
\end{equation}
Here, $V$ is the sample volume, $b_{\rm H} = 2.7 \times 10^{-15} \, \mathrm{m}/\mu_{\rm B} = 2.9 \times 10^{8} \, \mathrm{A}^{-1} \mathrm{m}^{-1}$ (with $\mu_{\rm B}$ the Bohr magneton), $\widetilde{\mathbf{M}}=\left[\widetilde{M}_x(\textbf{q}),\widetilde{M}_y(\textbf{q}),\widetilde{M}_z(\textbf{q})\right]$ denote the Fourier coefficients of the magnetization in the $x$-, $y$- and $z$-directions, and $^*$ indicates the complex conjugate. The angle $\Theta$ is enclosed between the magnetic field $\mathbf{H}$ and the scattering vector $\mathbf{q}=(0,\sin \Theta,\cos \Theta)$.  Inelastic phonon and magnon scattering is suppressed at small scattering angles \cite{Hansen2000}. Spiral and nuclear-spin incoherent scattering terms are neglected in Eq.~(\ref{Eq1}), and consequently $d \Sigma^{+ -} / d \Omega = d \Sigma^{- +} / d \Omega$. 
The minimum field strength to provide a sufficient guide field and to maintain the polarization of the neutrons was $\mu_0H=2\,\mathrm{mT}$, and the maximum field strength we could apply with an electromagnet was $\mu_0H=1\,\mathrm{T}$.
At 2\,mT we can assume the particle moments are on average randomly oriented, whereas at 1\,T we are close to saturation (see Fig.\ref{Fig1}(b)). In the following, we will present the results of these two extreme cases.

\subsection{Magnetic SANS close to saturation}
Close to magnetic saturation, the transversal magnetization components are suppressed (i.e., the real space magnetization $M_{x,y} \ll M_z\approx M_{\rm S}$), and the $d \Sigma^{+ -} / d \Omega$ shows a $\mathrm{sin}^2\Theta\mathrm{cos}^2\Theta$ anisotropy. In Fig.~\ref{Fig2}(a) we plot $d \Sigma^{+ -} / d \Omega(q,\Theta)$ detected at 1\,T (composite of the two patterns detected at the two detector distances) and Fig.~\ref{Fig2}(b) shows the normalized radial average $d \Sigma^{+ -} / d \Omega(\Theta)$. The functional form is proportional to $\widetilde{M}_z^2 \sin^2\Theta\cos^2\Theta$ plus a background, which shows that at 1\,T the moment ensemble is preferentially, but not completely, aligned along $z$-direction. This agrees with the isothermal magnetization curve from Fig.~\ref{Fig1}(b), which was close to magnetic saturation at 1\,T with a normalized magnetization of 0.966. From the integral magnetization measurement, we can estimate that the  magnetic moments are on average misaligned by $15 ^{\circ}$ with respect to the magnetic field, i.e., the absence of a magnetically saturated microstructure. The transversal magnetization components give rise to spin-misalignment scattering and are caused by either (i) variations in strength and orientation of a randomly perturbing field (e.g., magnetocrystalline anisotropy or thermal fluctuations) from particle to particle, or (ii) the dipolar stray fields generated by the neighboring particles \cite{Honecker2013Theory}.

For the approach to saturation at 1\,T, we  will now analyze the magnetic microstructure in the powder using micromagnetics. The sample can be considered as a space-filling array of interacting nanometer-sized, single-domain grains with random orientation of easy axes. The scattering features close to magnetic saturation can be analyzed with micromagnetic theory for the SANS of two-phase magnets \cite{Honecker2013Theory}.  In the limit of small misalignment of the magnetic moments from the mean magnetization, the micromagnetic balance of torque can be linearized and an analytical solution of the Fourier coefficients of the magnetization vector field are found.  This approximation involves that we assume $M_z \cong M_{\rm S}$  in real space. The approach provides closed-form expressions for the magnetization vector field, and  allows to evaluate the magnetic SANS and extract characteristic magnetic parameters like exchange interaction, strength of the magnetostatic stray field, or magnetic anisotropy field. For a self-contained discussion, we provide here the basics of the micromagnetic SANS description.
The demagnetizing field $\mathbf{H}_d(\mathbf{r})$ is determined by Maxwell's equations in the absence of any current. In Fourier space, the Fourier coefficient $\mathbf{h}_d(\mathbf{q})$ of $\mathbf{H}_d(\mathbf{r})$ is given by 
\begin{equation}
\label{hdbdef}
\mathbf{h}_d(\mathbf{q}) = - \frac{\mathbf{q} \, [\mathbf{q} \cdot \mathbf{\widetilde{M}}(\mathbf{q})]}{q^2} \,,
\end{equation}
which offers a convenient method to compute the magnetostatic stray field 
\begin{equation}
\label{hdbfourierdef}
\mathbf{H}_d(\mathbf{r}) = \frac{1}{(2\pi)^{3/2}} \int \mathbf{h}_d(\mathbf{q}) \, \exp(i \mathbf{q} \mathbf{r}) \, d^3q \,.
\end{equation}
Likewise, we also introduce the Fourier transform $\mathbf{h}(\mathbf{q}) = (h_x(\mathbf{q}), h_y(\mathbf{q}), 0)$ of a random perturbation $\mathbf{H}_p(\mathbf{r})$, e.g., the magnetocrystalline anisotropy field or a randomly fluctuating thermal field,  as
\begin{equation}
\label{Hpfourierdef}
\mathbf{H}_p(\mathbf{r}) = \frac{1}{(2\pi)^{3/2}} \int \mathbf{h}(\mathbf{q}) \, \exp(i \mathbf{q} \mathbf{r}) \, d^3q \,,
\end{equation}
which reflects the details of the microstructure (e.g., grain size and  crystallographic texture).
One obtains for a general orientation of the wave vector $\mathbf{q} = (q_x, q_y, q_z)$  the small-misalignment solution for the transversal Fourier coefficient $\widetilde{M}_x(\mathbf{q})$ and $\widetilde{M}_y(\mathbf{q})$. For the perpendicular scattering geometry ($\mathbf{H}\perp \mathbf{k}_0$), implicitly an average over the sample thickness along the beam path is performed (i.e., $q_x= 0$ in Fourier space) and the Fourier coefficient of the magnetization vector field then simplify to 
\begin{eqnarray}
\label{solmxqx0}
\widetilde{M}_x(\mathbf{q}) &=& M_{\rm s} \, \frac{h_x(\mathbf{q})}{H_{\mathrm{eff}}} \,, \\
\label{solmyqx0}
\widetilde{M}_y(\mathbf{q}) &=& M_{\rm s} \, \frac{h_y(\mathbf{q}) - \widetilde{M}_z(\mathbf{q}) \frac{q_y q_z}{q^2}}{H_{\mathrm{eff}} + M_{\rm s} \frac{q^2_y}{q^2}} \,, \\
\label{solmzqx0}
\widetilde{M}_z(\mathbf{q}) &=& \frac{1}{(2\pi)^{3/2}} \int M_z(\mathbf{r}) \, \exp(- i \mathbf{q} \mathbf{r}) \, d^3r \,,
\end{eqnarray}
where $h_{ x}(\mathbf{q})$ and $h_{ y}(\mathbf{q})$ represent the Cartesian components of the Fourier transform of the perturbing field $\mathbf{H}_{ p}$, and $H_{\rm eff}(q, H) = H_i \left( 1 + l_{ H}^2 q^2 \right)$ with $l_{ H}(H) = \sqrt{2 A / (\mu_0 M_{\rm S} H_i)}$ denote, respectively, the effective magnetic field and the exchange length of the field ($M_{\rm S}$: saturation magnetization; $A$: exchange-stiffness parameter; $\mu_0$: permeability of free space). $H_i\cong H-N_d M_{\rm S}$ is the internal field, i.e., the applied magnetic field corrected for the demagnetizing field, where  $N_d$ denotes the demagnetizing factor.
The Fourier coefficient of the longitudinal magnetization $\widetilde{M_z}$ follows from jumps of the magnetization magnitude over phase boundaries or particle matrix interfaces. The transversal magnetization component along the beam direction $\widetilde{M}_x$ (Eq.~\ref{solmxqx0}) only depends on the perturbing field provided by the individual particles, i.e., the expressions has become insensitive to the demagnetization field. The other transversal magnetization component $\widetilde{M}_y$ contains  in the numerator a term probing the longitudinal magnetization fluctuations $\widetilde{M}_z$ and in the denominator a term $M_{\rm S} \sin^2 \Theta$. Both are signatures the pole-avoidance principle of magnetostatics \cite{Perigo2014} and reflect the highly anisotropic dipole-field-type spin texture decorating the dipolar stray field. 
By inserting the analytical solutions for the Fourier amplitudes of the magnetization vector field  in Eq.~\ref{Eq1}), we can express the cross section near magnetic saturation of an ensemble of magnetic nanoparticles with a random distribution of magnetic easy axis as
\begin{equation}
\label{sigmasmperpsf}
\frac{d \Sigma^{+-}}{d \Omega} = \frac{8 \pi^3 b_H^2}{V}  \left( h^2(q) \, R_H + \widetilde{M}_z^2(q) \, R_M \right) \,,
\end{equation}
with the micromagnetic response functions
\begin{equation}
\label{rhdefperpsf}
R_H(q, \Theta, H)= \frac{p^2}{2} \left( 1 + \frac{\cos^4\Theta}{\left( 1 + p \sin^2\Theta \right)^2} \right) \,,
\end{equation}
and
\begin{equation}
\label{rmdefperpsf}
R_M(q, \Theta, H)=\sin^2\Theta \cos^2\Theta \left(1+  \frac{ p \,  \cos^2\Theta}{1 + p \sin^2\Theta}\right)^2 \,,
\end{equation}
where $p(q, H) = M_{\rm S}/H_{\rm eff}$. The scattering cross-section depends hence on a contribution related to the Fourier coefficient $h(\mathbf{q})$ of the perturbing field, and a contribution related to  $\widetilde{M}_z(\mathbf{q})$, which is associated to magnetostatic fields originating from the divergence of the magnetization at particle interfaces. Neglecting direct exchange interaction  --- absent between particles separated by an organic surfactant  --- implies $A=0$ for the macrospin arrangement   the effective field simplifies to $H_{\rm eff}= H_i$. The first term in the response function (Eq.~\ref{rmdefperpsf}) is the scattering contribution one would observe for a perfectly saturated spin system, and the second term is related to additional spin-misalignment scattering.

We applied the above theory, which was originally developed for bulk ferromagnets, to analyze the data of the nanoparticle powder close to saturation.
First, we fitted the azimuthal average of the scattering intensity within the $q$-range of nearest neighbor correlations in Fig.~\ref{Fig2}(b) with a weighted superposition of $R_{\rm H}$ and $R_{\rm M}$.
Although the fit is not perfect (which may be explained by additional energy contributions not considered in the micromagnetic theory, such as stemming from surface anisotropy \cite{gazeau1998,yanes2007,perez2008}), it becomes evident that the term $R_{\rm H}$ is needed to account for the increased scattering along the field direction compared to the vertical direction.
This indicates that the spin misalignment at high field is partially caused by a randomly oriented perturbing field, like thermal noise.
Fitting of the experimental 2D data [Fig.~\ref{Fig2}(a)] to Eq.~\ref{sigmasmperpsf} allows to extract the Fourier transform of the perturbing field $h^2(q)$ and of the magnetization in field direction $\widetilde{M}_z^2(q)$, which are both plotted in Fig.~\ref{Fig2}(c). 
In Fig.~\ref{Fig2}(c) we plot additionally the purely nuclear cross-section, which we obtained from the non-spin-flip cross-section, and it is easily seen that its low-$q$ behavior significantly deviates from $\widetilde{M}_z^2$. 
Nevertheless, we observe also for $\widetilde{M}_z^2$  an increase for the smallest momentum transfer vector $q<0.2\,$nm, which is a clear signature of positive, interparticle correlations (i.e., parallel alignment) of the neighboring particle moments at high fields. 
Regarding  $h^2(q)$, we would expect for an ideal, random system, that the perturbing field follows the single particle form factor, which exhibits a Guinier plateau at low $q$ range \cite{Michels2020} as shown in Fig.~\ref{Fig2}(c).
However, the upturn of the Fourier coefficient  $h^2(q)$ of the random-perturbation field at small $q$ indicates a non-zero effective magnetocrystalline anisotropy of the cluster (preferred local orientation), or, alternatively, collective thermal excitations \cite{Tartakovskaya2008,Krycka2018}.
Finally, it can be noted that $\widetilde{M}_z^2(q)$ and $h^2(q)$ are of the same magnitude, which demonstrates that the fluctuations in the longitudinal magnetization have comparable strength as the random-perturbing field of the crystallites, and hence the same influence on the magnetic energy of the system.  

\subsection{Magnetic SANS in the remanent state}
Fig.~\ref{Fig2}(d) shows the scattering pattern $d \Sigma^{+ -} / d \Omega(q,\Theta)$ observed for $\mu_0H=2\,\mathrm{mT}$ and Fig.~\ref{Fig2}(e) displays the azimuthal $d \Sigma^{+ -} / d \Omega(\Theta)$.
The scattering anisotropy obeys a $(\widetilde{M}_x^2+\widetilde{M}_{\perp}^2\cos^2\Theta)$ behavior indicating a random distribution of the particle moments with $\widetilde{M}^2_y=\widetilde{M}_z^2=\widetilde{M}_{\perp}^2$ and $\widetilde{M}_x^2=\widetilde{M}_{\perp}^2$ for $q>0.3\,\mathrm{nm^{-1}}$, i.e., over {\it intra}particle and nearest neighbor distances. The cross term in Eq.~\ref{Eq1}, uneven in the trigonometric functions and linear in the longitudinal and transversal magnetization, is negligible. This agrees well with the isothermal magnetization curve [Fig.~\ref{Fig1}(b)] with $M_z/M_{\rm S}\approx0$ at $\mu_0H=2\,\mathrm{mT}$. For an ideal superparamagnetic ensemble in zero field, i.e., in the absence of correlations between the randomly oriented particle moments, the scattering cross-section would simply reflect the particle form factor assuming a homogeneous magnetization within the individual spherical particles \cite{wiedenmann2005}.  In a real system,  fluctuating spin-pair correlations between core moments can persist to temperatures above the superparamagnetic blocking temperature.  The elastic neutron scattering measures static, frozen correlations in a spin system, with the dynamics taking place on time scales larger than the neutron interaction time \cite{Felber1998}. Analysis of the radially averaged, 1D  sf data at low field suggest the presence of directional correlations between nearest neighbor moments with a clear preference for an short-ranged antiferromagnetic-(AFM)-like interaction among the superparamagnetic grains \cite{bender2018dipolar}.

A further, in-depth analysis of the 2D magnetic SANS cross-section allows to extract the magnetization Fourier components. As shown in Fig.~\ref{Fig2}(f), the squared Fourier coefficient $\widetilde{M}_x^2$ of the magnetization along the beam direction deviates from the particle form factor and exhibits a pronounced peak at around $q=0.12\,\mathrm{nm^{-1}}$.
The decrease for $q\rightarrow 0$ is explained by anticorrelations between neighboring particle moments, i.e., a tendency for an antiferromagnetic-like alignment, due to dipolar interactions \cite{bender2018dipolar}. 
We plot in Fig.~\ref{Fig2}(f) additionally the square magnetization component in the scattering plane $\widetilde{M}_{\perp}^2$, which has a different $q$-dependence, and does not decrease for $q\rightarrow 0$. Considering that the  particle moments are globally randomly distributed this discrepancy can be only explained by a directional dependence of the magnetic structure factor, i.e., a different functional form for the in-beam magnetization component compared to in-scattering-plane components.

Thus, our experimental data demonstrate two striking insights regarding the magnetic structure factor of the interacting particle ensemble, namely: (i) the structure factor is anisotropic in zero field, and (ii) even in saturation the magnetic structure factor deviates from the nuclear one.
In the following, we will derive theoretical expressions for the magnetic structure factor to explain our findings.

\section{the magnetic structure factor}
 The magnetic interference scattering is determined and reflects the anisotropic  Halpern-Johnson vector (or magnetic scattering vector) $\mathbf{Q}$ originating from the dipolar neutron spin-magnetization interaction. Similarly, as for the magnetostatics in the micromagnetic description, unequal magnetization components can be expected.
To further elucidate the variation of the magnetic structure factor with direction  and the discrepancy to the nuclear structure factor, we introduce spin-pair correlation functions for SANS. We focus on the sf scattering cross-sections, but the same approach can be easily extended to the non-spin-flip, half-polarized and unpolarized scattering cross-sections. The sf scattering cross-section due to an arrangement of magnetic moments (atomic or particle moments) can be generally written as \cite{moon69}
\begin{equation}
\label{mrk}
\frac{d\Sigma^{+-}}{d\Omega}=\frac{b_{\rm H}^2}{V}\sum_{i,j} \exp(-\mathrm{i}\mathbf{q r})\times [Q_{x,i}Q^{\ast}_{x,j}+Q_{y,i}Q^{\ast}_{y,j}]\,,
\end{equation}
 where $V$ denotes the scattering volume,  $b_{\rm H}=2.91\times 10^8 \mathrm{A}^{-1}\mathrm{m}^{-1}$ is the atomic magnetic scattering length and the distance vector $\mathbf{r}=\mathbf{r}_i-\mathbf{r}_j$ between two scattering entities. For the perpendicular scattering geometry  ($\mathbf{H}\perp\mathbf{k}_0$), the magnetic scattering vector equals to 
 \begin{equation}
 \label{HPvector}
 \mathbf{Q}=\frac{\mathbf{q}}{q^2}(\mathbf{q}\cdot \widetilde{\mathbf{M}})-\widetilde{\mathbf{M}}=
 \left(  \begin{array}{c}
 -\widetilde{M}_x \\
\widetilde{M}_z  \cos\Theta  \sin\Theta -\widetilde{M}_y  \cos^2\Theta \\
\widetilde{M}_y  \cos\Theta  \sin\Theta -\widetilde{M}_z  \sin^2\Theta \\ 
\end{array}
  \right)
 \end{equation}
The scalar product of the magnetic scattering vector is given by
 \begin{equation}
 \label{unpolCS}
\frac{d \Sigma}{d \Omega}(\mathbf{q}) \propto\mathbf{Q}\cdot\mathbf{Q}=\widetilde{\mathbf{M}}^2-\frac{|\widetilde{\mathbf{M}}(\mathbf{q})\cdot \mathbf{q}|^2}{|\mathbf{q}|^2}\,,
 \end{equation}
 which readily corresponds to the unpolarized magnetic scattering cross-section. Alternatively, one can express the differential elastic scattering cross-section by the Fourier transform of the pair correlation function of the magnetization component $\mathbf{M}_{\perp}$ perpendicular to the scattering vector $\mathbf{q}$ \cite{Lermer1976}.
The first term in Eq.~\ref{unpolCS} reflects the autocorrelation function of the magnetization, while the second term represents the product of magnetization with the magnetostatic field defining the magnetostatic energy 
 \begin{equation}
 \label{magstatEnergy}
E_M=\int \mathbf{H}_d(\mathbf{r})\cdot \mathbf{M}(\mathbf{r}) d^3 \mathbf{r}=\int \frac{ |\widetilde{M}(\mathbf{q})\cdot \mathbf{q}|^2}{|\mathbf{q}|^2} d^3 \mathbf{q}\,,
\end{equation}
 which reflects the dipolar interaction between the neutron spin and the magnetic field inside the sample\cite{HalpernJohnson1939}.

\begin{figure}
\centering
\includegraphics[width=0.8\columnwidth]{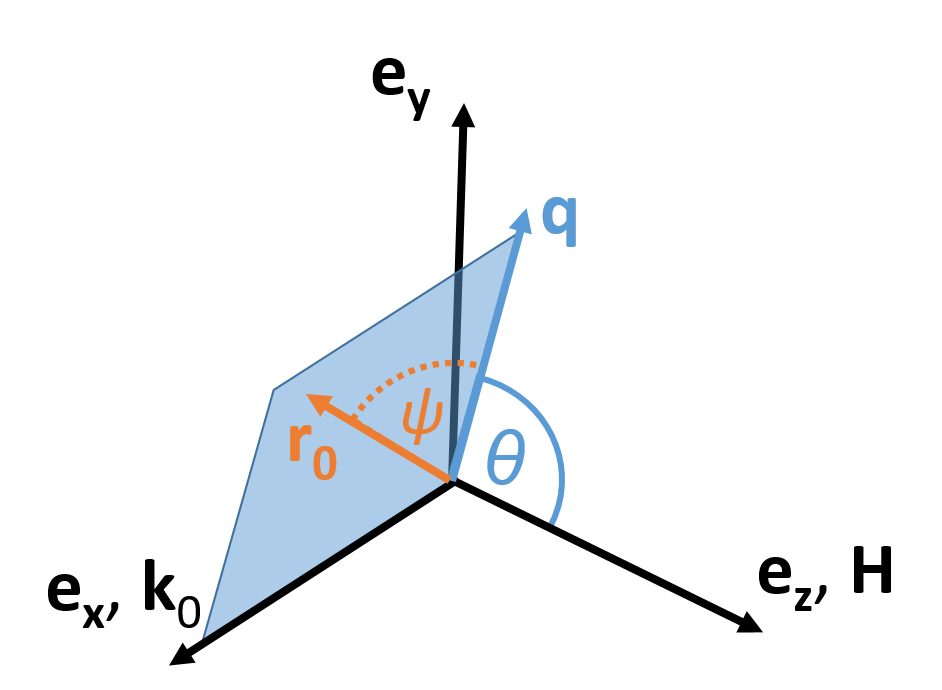}
\caption{\label{Fig3} The angle $\Theta$ denotes the orientation of the momentum transfer vector $\mathbf{q}$ in the detector plane spanned by $\mathbf{e}_y$-$\mathbf{e}_z$ and the angle $\Psi$ defines the orientation of the distance vector $\mathbf{r}$ relative to $\mathbf{q}$ and away from the scattering plane.}
\end{figure}

Here, we consider a disordered system of particles such that we can restrict the discussion to two-particle interference  \cite{Hove54}, but already the knowledge of the two-particle configuration is satisfactory to understand the main scattering features. The special case of three-particle correlations will be further discussed exemplary for the specific situation of cubic order.

Let us first consider the magnetization field of two particles with a uniform internal magnetization. The magnetization vector can be expressed with $M_{\rm P}$ the saturation magnetization, $\hat{\mathbf{m}}$ the unit vector of the magnetization direction and  the particle form factor $F(\mathbf{q})$ assumed equal for the involved particles. The Fourier transform of the magnetization for identical centro-symmetric particles is given as
  \begin{equation}
 \label{totalMrec}
 \widetilde{\mathbf{M}}(\mathbf{q})=\frac{1}{(2\pi)^{3/2}}\sum_{i=1}^2 M_{\mathrm{P}} V_{\mathrm{P}}\hat{\mathbf{m}}\, F(q)\exp(-\mathrm{i}\mathbf{\mathbf{q} \mathbf{r}_i})\,,
 \end{equation}
with $V_{\rm P}$ the particle volume.
The  SANS signal is composed as the incoherent sum over individual coherence volume containing two-particle correlations.
Inserting  Eq.~\ref{totalMrec} in Eq.~\ref{HPvector}, and combining with Eq.~\ref{mrk}, we obtain 
 \begin{equation}
\label{CS_explicit}
\begin{aligned}
\frac{d\Sigma^{+-}}{d\Omega} &= n_{\rm P}  V_{\rm P}^2 F^2(q) \\
&\left(\eta_x^2 S_x+\eta_y^2 S_y \cos^4\Theta +\eta_z^2 S_z \sin^2\Theta \cos^2\Theta \right)\,,
\end{aligned}
\end{equation} 
with $i=\{x,y,z\}$, $n_P$ the particle number density, $\mathbf{\eta}=b_{\rm H} \Delta\mathbf{ M}$ the magnetic scattering length density which depends on the magnetic contrast $\Delta M_i$ between the particle and the average magnetisation of the medium.

For pair correlations, the magnetic structure factor over $N$ particles is given by
 \begin{equation}
\label{magS_explicit}
\begin{aligned}
S_i(\mathbf{q}) &= 1+\frac{1}{N} \sum_{k=1}^N \langle \hat{\mathbf{m}}(0)\hat{\mathbf{m}}(r_k)\rangle_i \times \\
&\sum_{l=0,2,4, \cdots}\mathrm{i}^l (2l+1)   j_l(q r_k)P_l(\cos \Psi_k)\,,
\end{aligned}
\end{equation}
with  $P_l$ denoting the Legendre polynomial and $j_l$ the spherical Bessel function of order $l$, $\Psi_k$  the angle between the scattering vector $\mathbf{q}$ and the connecting vector $\mathbf{r}_k$ of a particle pair. The scalar product $\langle \hat{\mathbf{m}}(0)\hat{\mathbf{m}}(r_k)\rangle_i$  gives the average alignment in the direction $i$ at a interparticle distance $r_k$, which is positive for ferromagnetic-like correlations and negative for antiferromagnetic-like correlations.

Thus, the second term of $S_i(\mathbf{q})$ describes the magnetic interparticle interference of two particles with a specific orientation of their magnetization and relative distance $\mathbf{r}$. The scalar product between scattering vector and distance vector is $\mathbf{q}\cdot \mathbf{r}=q\, r \cos\psi$ so that the phase factor in Eq.~\ref{mrk} simply reads $\exp(-\mathrm{i}\mathbf{q r})=\cos(q r \cos\Psi)$. This definition allows to determine the scattering cross-section observed on a 2D position sensitive detector as used in SANS.  It is important to understand, that in SANS actually not the 3D correlations are determined, but the two-dimensional projection with $q_x=0$ and hence an integration over the beam direction is implicitly performed \cite{fritz2013two,mettus2015small}. 
The diffuse scattering of spin pairs is expressed usually as a function of the scattering vector magnitude $q$, and the orientational average over random crystallographic axes is obtained by rotating $\mathbf{q}$ around a local coordinate system given by $\mathbf{r}$ \cite{paddison2013spinvert}. 
 In contrast, here we define the distance vector $\mathbf{r}$ with respect to $\mathbf{q}$ (see Fig.~\ref{Fig3}) as
\begin{equation}
\label{distancer}
\mathbf{r}=r\left(   \begin{array}{c}
 \sin\Psi  \\
 \cos\Psi \sin\Theta \\
 \cos\Psi \cos\Theta 
\end{array}\right)\,,
\end{equation}
with $\Psi$ being the tilt angle of $\mathbf{r}$ away from the detector plane. 
The vector quantities involved in the magnetic scattering cross-section can be then expressed in a local Cartesian coordinate frame spanned by basis vectors parallel to the distance vector $\mathbf{r}$ and given by the orthogonal vector  $\mathbf{q}\times \mathbf{r}$ as well as the component of the distance vector $\mathbf{r}_{\perp \mathbf{q}}=-\mathbf{q} \times \mathbf{q}\times \mathbf{r}$ perpendicular to $\mathbf{q}$. Expressing the scattering vector in terms of these coordinates, it is easily shown that only the contribution $\mathbf{q}_{\parallel r}= \mathbf{q}\cdot \mathbf{r}\, \mathbf{r} $ acts for magnetic scattering. Similarly the magnetic scattering vector $\mathbf{Q}$ can be decomposed into parallel and perpendicular components with respect to the distance vector $\mathbf{r}$ as it is done in the neutron diffraction literature to obtain magnetization components with respect to $\mathbf{r}$\cite{blech1964spin}. For transparency, we choose to represent here the magnetization in terms of their Cartesian representation with z being the applied magnetic field direction as it is the conventional notation for field-dependent magnetic SANS. Note that a simple coordinate transform has no influence on the result of a scalar product like the ones involved in the scattering cross-section, in particular the result of the scalar product ($\mathbf{M}\cdot \mathbf{q}$) are unaffected. It is also readily seen that for the component $\mathbf{Q}_{x}$ (Eq.~\ref{HPvector}) along the beam only the autocorrelations  of the magnetization component $\mathbf{M}_x$ in the scattering plane are probed. 

Now we perform the angular integration  $\frac{1}{2\pi}\int_0^{\pi} \frac{d\Sigma^{+-}}{d\Omega} \sin \Psi  d\Psi $ assuming an isotropic microstructure, i.e., equal probability to find a particle at an arbitrary orientation with respect to $\mathbf{q}$.
 Considering monodisperse, homogeneously magnetized (spherical) particles, for the magnetic scattering cross-section follows:
 \begin{equation}
\label{CS_avg}
\begin{aligned}
\frac{d\Sigma^{+-}}{d\Omega}&=n_{\rm P} \eta^2(H) V_{\rm P}^2 F^2(q) \left( S_0(q , \hat{\mathbf{m}}_x)+ \right. \\
&\left. S_2(q , \hat{\mathbf{m}}_y) \cos^4\Theta  + S_2(q , \hat{\mathbf{m}}_z) \sin^2\Theta \cos^2\Theta \right) \,,
\end{aligned}
\end{equation}
using a short-hand notation for the field-dependent magnetic scattering length density $\eta^2(H)$, which should be considered as a vector (compare Eq~\ref{CS_explicit}).
The  structure factor averages are given by
\begin{equation}\label{Eqs0}
S_n(q,\hat{\mathbf{m}}_i)=1+\int  t(r) \langle \hat{\mathbf{m}}(0)\hat{\mathbf{m}}(r)\rangle_i B_n(qr) dr\,,
\end{equation}
with $B_n(z)=(-1)^n \frac{d^{2 n} j_0(z)}{d z^{2 n}}$ denoting derivatives of the zeroth-order spherical Bessel function $j_0(z)=\mathrm{sin} z/z$.
The spatial pair correlation function $t(r)=4\pi r^2 \rho(r)$, also observed for nuclear scattering, is determined by the particle density $\rho(r)$ at a distance $r$ from a particle in the origin. As can be seen in Eq~\ref{CS_avg}, the structure factor for the in-scattering-plane magnetization components  $\widetilde{M}_y$ and $\widetilde{M}_z$ have a different functional form involving also higher-order spherical Bessel functions compared to the in-beam magnetization contribution.

For the perfect (super)paramagnetic case, ignoring possible local short-range order,  the value of the magnetic structure factor is unity (see Eq.~\ref{Eqs0}) and only the pure particle form factor is obtained for $\frac{d\Sigma^{+-}}{d\Omega}$. However, a magnetic structure factor may appear as soon as short-range interparticle magnetic correlations exist due to some coupling mechanism, e.g., dipole-dipole interaction. Magnetostatic stray fields and the ensuing magnetic coupling between particle moments can locally break the symmetry usually ascribed to a isotropic (super)paramagnetic state at zero magnetic field. The spin-spin correlations between nearest, dipolar-coupled neighbors do not vanish, however, the average moment of the sample can be still negligibly small. In this case, we obtain for a macroscopically isotropic ensemble ($\widetilde{M}^2_x=\widetilde{M}^2_y=\widetilde{M}^2_z$, and $\langle M \rangle =0$) the simplified scattering cross-section
  \begin{equation}
\label{CS_zerofield}
\frac{d\Sigma^{+-}}{d\Omega}_{H=0}=n_{\rm P} \eta_{H=0}^2 V_{\rm P}^2 F^2(q)( S_0(q, \hat{\mathbf{m}})+S_1(q, \hat{\mathbf{m}}) \cos^2\Theta )\,.
\end{equation}
 At zero field, the magnetic scattering length density contrast of a superparamagnetic particle amounts to $\eta_{H=0}^2=\frac{1}{3} b_{\rm H}^2 M_{\rm P}^2$. 

As one can easily imagine the magnetic correlations vary with magnetic field. They evolve from short-range anticorrelations at low fields to a co-alignment (i.e., positive correlations) of the neighboring particle moments at high fields. At magnetic saturation ($\widetilde{M}^2_x,\widetilde{M}^2_y=0$) the sf scattering cross-section is given by
\begin{equation}\label{CS_maxfield}
\frac{d\Sigma^{+-}}{d\Omega}_{\rm sat}=n_{\rm P} \eta_{\rm sat}^2 V_{\rm P}^2 F^2(q) S_2(q)\mathrm{sin}^2\Theta\mathrm{cos}^2\Theta\,,
\end{equation}
where the scattering length density  $\eta_{\rm sat}$ is the difference between the saturation-magnetization values $M_{\rm P}$ of the particle and the integral saturation magnetization $M_{\rm S}$ of the whole sample. The structure factor follows here $S_2(q)$, which has the substantial consequence that the usual assumption of nuclear and magnetic structure factor being identical is for spin-resolved measurements  never true.

The above calculations are valid for pair correlations, but higher-order correlations may arise depending on the local structural symmetries \cite{Altarelli2010}. As an example, we now consider the special case three particles forming an isosceles right triangle. The higher order structure factor would then have a modified form 
\begin{equation}\label{Smagthirdorder}
\begin{aligned}
S_n(q,\hat{\mathbf{m}}_i)&=1+\int  t(r) \langle \hat{\mathbf{m}}(0)\hat{\mathbf{m}}(r)\rangle_i \\
&[(1-b(r)) B_n(qr) +\frac{b(r)}{2} B_{n-1}(qr) ]dr\,,
\end{aligned}
\end{equation}
with  $b(r)$ the fraction of particles exhibiting orthogonal three particle correlations at a given distance $r$. Thus, $b(r)$ contains information on the local particle environment. The local symmetry parameter $b(r)$  varies between 0 for a disordered system (i.e., if no higher order structure exists) and unity if the orientation of the magnetization is not correlated with the direction of $\mathbf{r}$, i.e. for a microstructure exhibiting a homogeneous magnetic scattering length density (e.g., inside a particle or reaching the mean medium limit at large $r$) or at least a cubically ordered array of magnetic entities such that magnetic density correlations in orthogonal directions are equal. This is in line with simulations of the 3D magnetic correlations in frustrated magnets seen by magnetic diffuse neutron scattering, which show that for cubic structures only the autocorrelation function of the magnetization vector field persists whereas the higher order term cancel \cite{Roth2018}. For this special case of orthogonally ordered particles or homogeneous magnetic density, the argument under the integral for the magnetic structure factors converges to the commonly expected form $j_0(q r)=\mathrm{sin} (qr)/qr$, e.g., for the sf scattering cross-section at zero field, but also in the case of the unpolarized scattering cross-section at magnetic saturation. Eventually, for a perfectly aligned array of moments, magnetic and nuclear structure factor may then coincide. The anisotropy of the magnetic structure factor for a disordered microstructure ($S_1\neq S_0$ for $b(r)< 1$ in Eqs.~\ref{CS_zerofield} and \ref{CS_maxfield}), explains the observed strikingly different scattering behavior for  $|\widetilde{M}_x|^2$ and $|\widetilde{M}_y|^2$ at 2\,mT (Fig.~\ref{Fig2}(f) ) and discloses the significant deviation of $|\widetilde{M}_z|^2$  from the cross sections detected in low field as well as from the purely nuclear cross-section [comparing Fig.~\ref{Fig2}(c) and (f)].

Away from the two scenarios at zero and saturating magnetic field discussed here, further knowledge on the magnetization vector field is needed, e.g., from a (numerical) determination of the field dependent orientation function and the directional correlations between the core moments, to assess the diffuse neutron scattering.

\section{conclusion}

We present here expressions for the magnetic structure factor in SANS, which are suitable to investigate the local short-range order and magnetic correlations in clusters and structured assemblies of single-domain nanoparticles. 
The introduced formalism does not only describe the unpolarized scattering cross-section at zero field but also the spin-resolved, magnetic 2D SANS cross-section due to interparticle interference based on spin-pair correlations in a 3D spin ensemble. 
We demonstrate that the magnetic structure factor  is anisotropic for a structurally disordered ensemble (i.e., in absence of cubic symmetry).  
Our formalism is adequate to disclose in detail the influence of magnetic particle correlations with SANS.
Neglecting spin pair correlations, it is customary assumed for SANS that the magnetic structure factor (arising from moment correlations) is identical to the nuclear structure factor, which is given by the spatial particle arrangement and the concomitant pair distance correlation function. 
However, here we show that this assumption is not true in general and the magnetic structure factor can deviate significantly from the nuclear structure factor for magnetically interacting nanoparticle ensembles. 
This can lead to questionable findings and misinterpretations of the spatial magnetization profile, e.g., the postulation of an antiferromagnetic order or of a highly textured magnetic microstructure in an otherwise isotropic system based on SANS result, but which is not observed by complementary methods like integral magnetization measurements.

Our approach allows to explain the results of a polarized SANS experiment on a powder of dipolar-coupled 10-nm iron oxide nanoparticles. 
We verify that (i) at low field the structure factor along the beam direction deviates from the in-scattering-plane components and (ii) that close to saturation the magnetic structure factor along the field-direction deviates significantly from the nuclear structure factor.
The here developed spin-resolved framework may foster new insights using polarized neutrons and is generally applicable to analyze the magnetic pair distribution of local correlations, e.g. also with polarized neutron diffraction.

\acknowledgments
We thank M. Puerto Morales, R. Costo and H. Gavil\'an for the particle synthesis and the TEM analysis, E. Wetterskog for performing the magnetization measurements and D. Gonz\'alez-Alonso for his help during the SANS experiments. We also thank the Institut Laue-Langevin for provision of beamtime at the instrument D33. We would like to thank the anonymous referee, who pointed out a shortcoming in the first draft and suggested a extended revision of the manuscript.
This project has received funding from the European Commission Framework Programme 7 under grant agreement no 604448 (NanoMag). 
Philipp Bender acknowledges financial support from the National Research Fund of Luxembourg (CORE SANS4NCC grant).

\bibliography{PBenderBib}

\end{document}